# A Dynamic Interference Model for Benham's Top

Yutaka Nishiyama
Osaka University of Economics,
2, Osumi Higashiyodogawa Osaka, 533-8533, Japan
http://yutaka-nishiyama.sakura.ne.jp/index.html
nishiyama@osaka-ue.ac.jp

**Abstract**

In 2023, I published a paper titled "A Dynamic Interference Model for Benham's Top." Here, I would like to concisely explain the most important aspect of dynamic interference by organizing and integrating the figures.

## 1. Introduction

Research on subjective colors in perceptual psychology began with Prevost (1826) and continued with Fechner (1839) and Benham (1894) [1]. Charles Benham (1860–1929) invented Benham's top, which creates the appearance of colors from black and white patterns. This toy became hugely popular in Britain and was even featured in *Nature*, but the reason for its effect remains unknown to this day.

In 1979, I hypothesized that some form of interference that leads to the appearance of subjective colors might be at work in Benham's top [2]. While 45 years have passed since then, I have recently been able to complete this hypothesis [3]. Specifically, using mathematical and physical principles, I would like to explain how two consecutive light stimuli create dynamic interference when processed through the human visual system, and how subjective colors appear in order of wavelength.

## 2. From Polar to Cartesian Coordinates

Let's examine Benham's top. It's a disk approximately 10 cm in diameter, with the lower semicircle filled in black and the upper semicircle containing three arcs each (Fig. 1-1). The arcs are arranged in four blocks, each spanning 45 deg. While there could be two arcs or even one, three arcs produce the most easily visible effect. When the top spins clockwise (rightward), colors appear in the order of blue, green, orange, and red from the outside. These colors, while not vivid, are clearly perceivable as colors rather than black or white [4]. When spun counterclockwise (leftward), the colors appear in reverse order: red, orange, green, then blue. Two key insights for understanding these subjective colors are that reversing the spinning direction reverses the color order, and that the colors appear in the same wavelength order as a rainbow.

For theoretical development, I shifted the position of the outer and inner arcs by about 10 center-angular degrees, as shown by the arrows in Fig. 1-2, to prevent them from touching the black lower semicircle. This modification made the subjective colors, which had appeared somewhat dark, more visible. As a result, all four blocks now follow a black–white–black–white pattern.

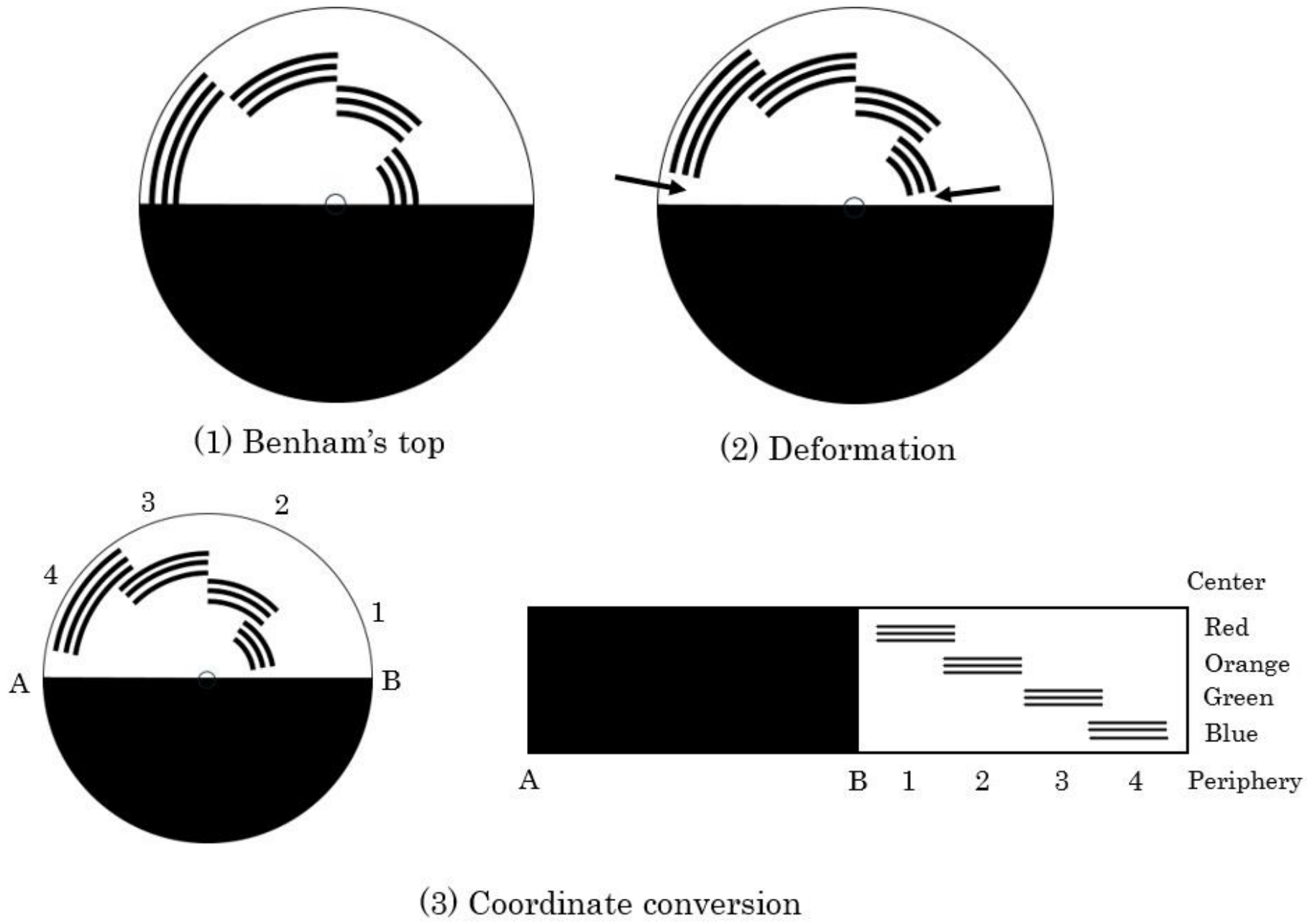

Fig. 1: From polar to Cartesian coordinates.

Converting from polar to Cartesian coordinates yields Fig. 1-3. Moving counterclockwise around the top's circumference, we label points A and B and numbers 1 through 4. Assume the disk is spinning rightward (clockwise). Fixing our viewpoint at point A on the left edge, the black portion from A to B passes first. The black lower half of the disk allows the eyes to rest and is unrelated to the subjective colors. After half a rotation, arc 1 appears, followed by arcs 2, 3, and 4. From the disk's center, subjective colors appear as red, orange, green, and blue. In Cartesian coordinates, these colors appear from top to bottom as red, orange, green, and blue.

All patterns from red to blue follow the black–white–black–white sequence. While the lengths of the first and second black segments are identical, the lengths of the white segments before and after the second black differ. If we consider the first white segment as the primary stimulus and the second white segment as the secondary stimulus, it is possible that these two stimuli create some form of interference (enhancement, cancellation, etc.).

## 3. Wave Superposition

Benham's top rotates 2–5 times per second, and because this pattern repeats, it can be regarded as an irregular rectangular digital wave. Cones in the retina absorb the light from the first white stimulus after it passes through the eyeball, and the brain recognizes it as white. Similarly, cones in the retina absorb light from the second white stimulus after it passes through the eyeball and retina, and the brain recognizes it as white. When white overlaps with white, it appears to remain white, seemingly leaving no room for colors to emerge.

The basic unit of biological information processing in the nervous system is the neuron (nerve cell). Information transmission between neurons occurs via ion conduction, which has characteristics such as a threshold (where pulses are not generated unless a stimulus exceeds a certain value) and delay (where transmission occurs not immediately but after a short period of time). I believe these characteristics may be related to the perception of subjective colors.

Now, let's consider the superposition of two sine waves with the same wavelength and amplitude, but with a phase difference α (Fig. 2):

$$y_1 = \sin\theta,$$
$$y_2 = \sin(\theta + \alpha).$$

The sine waves of the primary and secondary stimuli have a period of 5 (Fig. 2, top). When the phases overlap with a 90° shift ($\alpha = \frac{\pi}{2}$), the maximum amplitude of the wave becomes 1.4 times larger. With a 180° ($\alpha = \pi$) shift, the waves cancel each other out, and the amplitude becomes 0. With a 360° ($\alpha = 2\pi$) shift, the amplitude of the wave doubles (Fig. 2, bottom). The superposition of two sine waves, according to the trigonometric sum-to-product formula, becomes

$$y_1 + y_2 = \sin\theta + \sin(\theta + \alpha) = 2\cos\frac{\alpha}{2}\sin\left(\theta + \frac{\alpha}{2}\right).$$

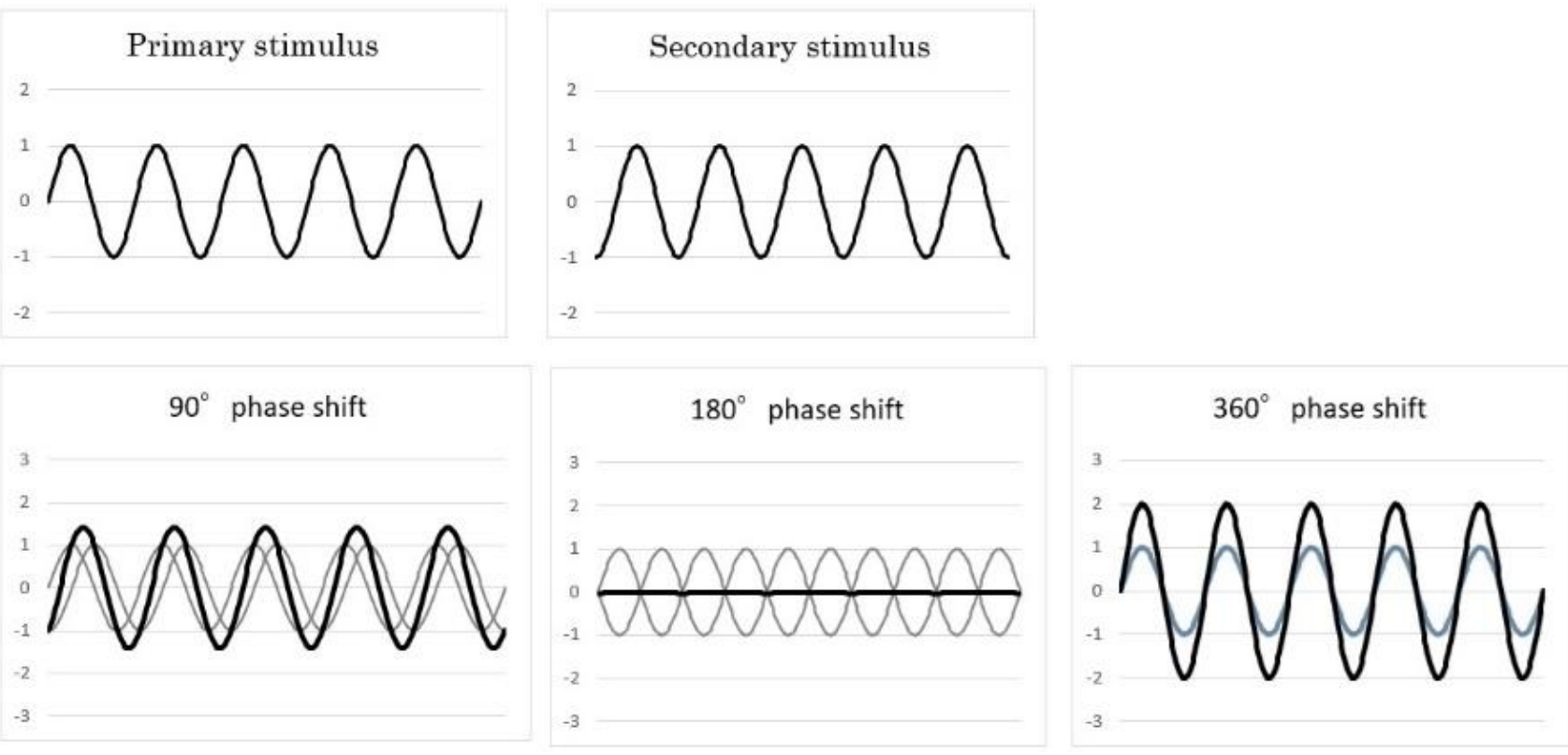

Fig. 2: Superposition of waves.

## 4. Dynamic Interference

When there is no phase difference (α = 0) between the primary and secondary stimulus waves, no interference occurs, and white overlapped with white remains white, with no colors emerging. When there is a phase difference between the two waves (α ≠ 0), however, interference occurs and specific colors emerge. While white light contains all colors, for simplicity, we'll consider just three colors: red, green, and blue (RGB).

Figure 3, though somewhat complex, consists of 16 graphs arranged in 4 rows and 4 columns. We set the period for red (R) to 4, green (G) to 5, and blue (B) to 6.

The first row represents red (R), showing graphs of phase-shifted red waves (R1), phase-shifted green waves (R2), and phase-shifted blue waves (R3) superimposed with the original red (R). R+R1 shows double amplitude, R+R2 shows 1.6 times amplitude, and R+R3 shows equal amplitude. When shifted by the red wavelength, red becomes prominently visible.

The second row represents green (G), showing graphs of phase-shifted red waves (G1), phase-shifted green waves (G2), and phase-shifted blue waves (G3) superimposed with the original green (G). G+G1 shows 1.4 times amplitude, G+G2 shows double amplitude, and G+G3 shows 1.7 times amplitude. When shifted by the green wavelength, green becomes prominently visible.

The third row represents blue (B), showing graphs of phase-shifted red waves (B1), phase-shifted green waves (B2), and phase-shifted blue waves (B3) superimposed with the original blue (B). B+B1 shows zero amplitude, B+B2 shows 1.6 times amplitude, and B+B3 shows double amplitude. When shifted by the blue wavelength, blue becomes prominently visible.

The fourth row shows the combination of all three colors (RGB). In RGB notation, the leftmost is (255, 255, 255) representing white. The second column is (255, 179, 0) showing ochre, the third column is (204, 255, 204) showing light green, and the fourth column is (128, 217, 255) showing light blue. The subjective colors keep red, green, and blue as their base while also retaining other colors, resulting in pastel tones.

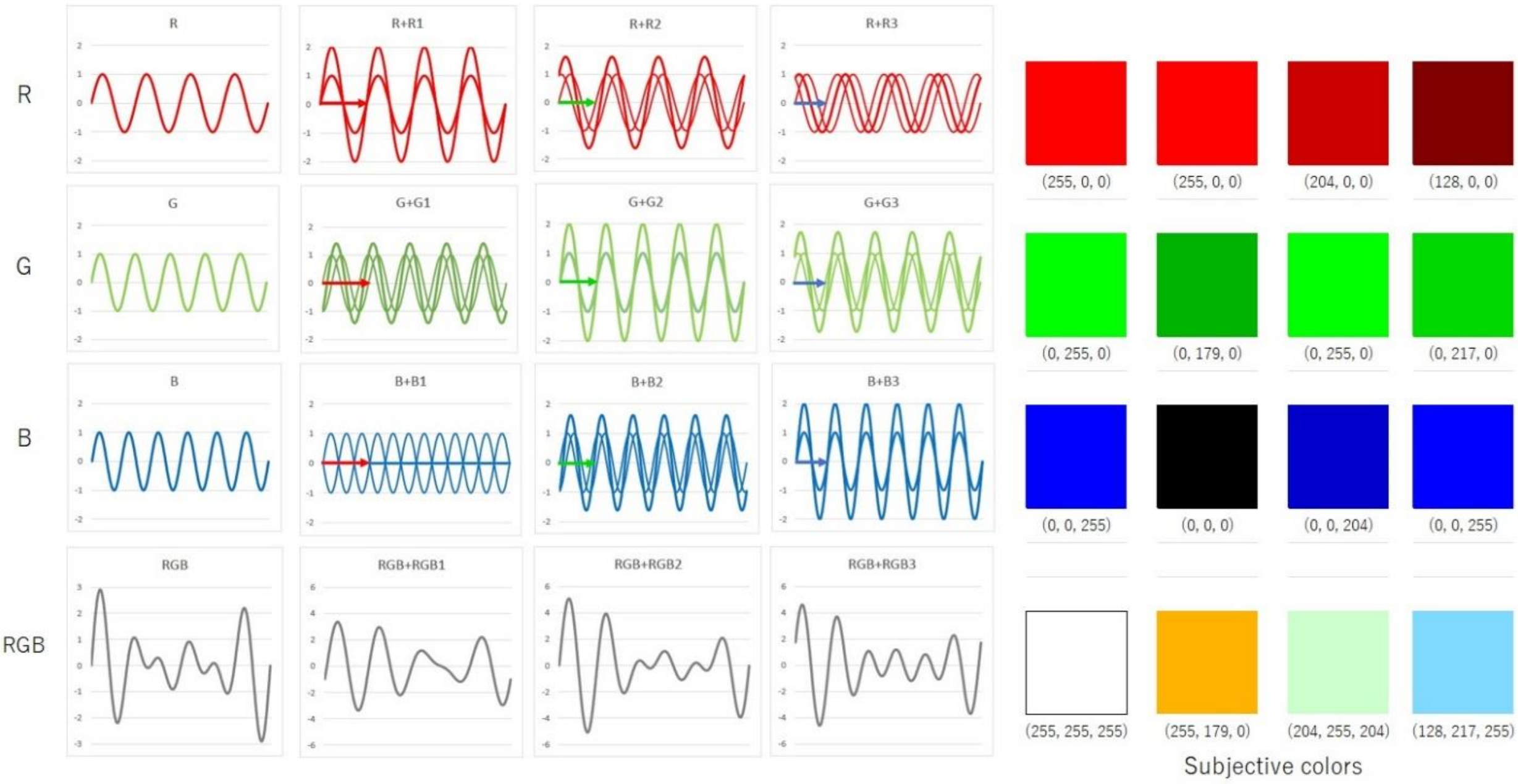

Fig. 3: Why certain colors emerge.

## 5. Equation of Motion for Delay

Let's consider how this corresponds to Benham's top. Figure 1-3 showed the rotating disk converted to Cartesian coordinates, which can be simplified as shown in Fig. 4. As the top rotates, it repeatedly progresses from left to right in Cartesian coordinates. From top to bottom, this corresponds to red, orange, green, and blue, with the wavelengths arranged from longest to shortest. Each case follows a black–white–black–white pattern, where the first black serves to rest the eyes and is unrelated to the subjective colors, while the subsequent white–black–white sequence is related. Let's refer to the first white as the primary stimulus and the next white after the black as the secondary stimulus. When the primary stimulus shifts and overlaps with the secondary stimulus, dynamic interference occurs (Fig. 4-1).

If we make the primary stimulus shift largest for red and smallest for blue, this corresponds to why specific colors emerge as shown in Fig. 3. But can such a convenient hypothesis hold true? Could the black between the white–black–white sequence delay and inhibit transmission of the primary stimulus? If we assume the delay is greater when the white interval of the primary stimulus is short and smaller when long, this would conveniently explain the subjective colors. I thus propose a bold hypothesis.

Let's consider the primary white stimulus as a quantity of light, transmitting substance, or moving object, corresponding to mass $m$ $(m_i)$ and acceleration $a$ $(a_i)$. If we regard the white area as a physical quantity that transmits white light to the visual system, we could analogize its magnitude to mass. When the white area is large, the light quantity is large and thus the mass is large; when the area is small, the light quantity is small and thus the mass is small. The black portions contain no white light and transmit nothing to the visual system. Rather, they might function to stop or delay the progression of the physical quantity that the preceding white has transmitted to the visual system.

Black acts as a force that delays the transmission of white. Since the length of black is the same in all cases, if we consider it as a force $F$ (constant), these can be expressed by the equation of motion

$$F = ma = m_i a_i.$$

From Fig. 4-2, since the length of the primary stimulus is shortest for red and longest for blue, the masses $m$ have the relation

$$m_1 < m_2 < m_3 < m_4.$$

From the equation of motion $F = ma$, since $F$ is constant, the accelerations $a$ become

$$a_1 > a_2 > a_3 > a_4.$$

So Newton's equation of motion holds true even in Benham's top: the smaller the interval of the primary stimulus, the greater the shift, and the larger the interval, the smaller the shift when there is overlap with the secondary stimulus, creating dynamic interference.

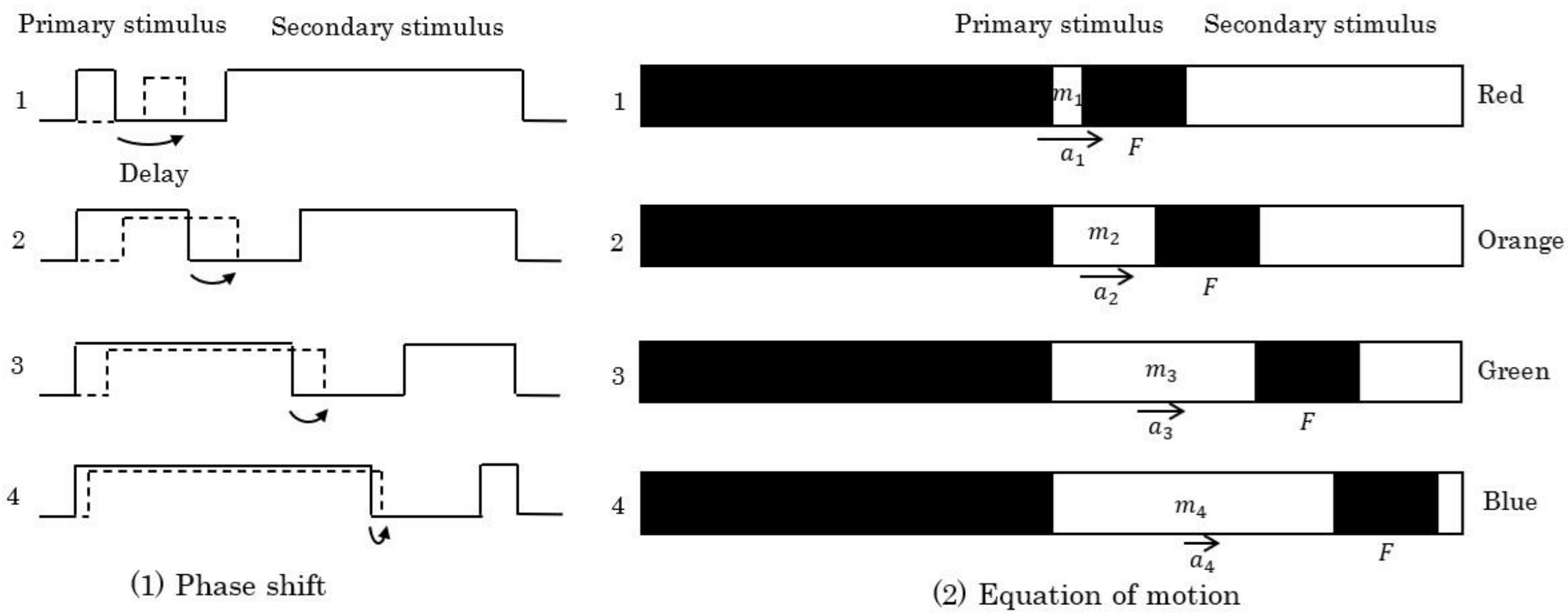

Fig. 4: Phase shift and equation of motion ($F = ma$) for delay.